\documentclass[twocolumn,aps,floats,preprintnumbers]{revtex4}
\usepackage{epsfig,amsmath,amsfonts,amssymb}
\begin{document}

\preprint{MZ-TH/08-07}
\preprint{February 29, 2008}

\title{Cosmon Lumps and Horizonless Black Holes}

\author{Timm Kr\"uger\,$^a$, Matthias Neubert\,$^b$\footnote{On leave from Laboratory for Elementary-Particle Physics, Cornell University, Ithaca, NY 14853, U.S.A.}, and Christof Wetterich\,$^a$
\vspace{0.2cm}}

\affiliation{$^a$ Institut f\"ur Theoretische Physik\\
Universit\"at Heidelberg\\
Philosophenweg 16, D-69120 Heidelberg, Germany\\ 
\\
$^b$ Institut f\"ur Physik (THEP)\\
Johannes Gutenberg-Universit\"at Mainz\\
D-55099 Mainz, Germany}

\begin{abstract}
We investigate non-linear, spherically symmetric solutions to the coupled system of a quintessence field and Einstein gravity. In the presence of a scalar potential, we find regular solutions that to an outside observer very closely resemble Schwarzschild black holes. However, these cosmon lumps have neither a horizon nor a central singularity. A stability analysis reveals that our static solutions are dynamically unstable. It remains an open question whether analogous stable solutions exist.
\end{abstract}

\maketitle

\section{Introduction}

Black holes are fascinating and intriguing objects, both by the presence of a horizon at the Schwarzschild radius $R_S$ and by the existence of a central singularity. They arise as solutions of pure gravity, and one may wonder if the characteristic black-hole properties can be influenced by the presence of other fields. In fact, adding a massless, free scalar field to gravity, new spherically symmetric solutions to the field equations have been found, for which the horizon is smoothed out \cite{Buchdahl:1959nk,Brans:1961sx,Wyman:1981bd} (see also \cite{Wetterich:2001cn}). In the exterior region (for $r>R_S$) these solutions can be made arbitrarily close to the Schwarzschild solution; however, the event horizon is replaced by a high red-shift horizon, at which all metric components remain finite. It was proposed that these ``approximate black holes" might be used in numerical simulations of black holes as a way of regularizing the the horizon \cite{vanPutten:1996mt}. Unfortunately, for free scalar fields the solutions exhibit a naked central singularity and were found to be unstable \cite{Jetzer:1992ag,Liebling:1996dx,Choptuik:1997rt}. In this Letter, we present spherically symmetric solutions for an interacting scalar field that are completely regular. For a suitable scalar potential, they look from the outside very similar to a Schwarzschild black hole, but neither an event horizon nor a central singularity exist. Observational distinction from black holes seems extremely difficult for many such solutions.

In order to influence the properties of the horizon, the additional field must be very light, with typical mass smaller than the inverse Schwarzschild radius. The proposal of a solution to the cosmological constant problem by a time-varying dark energy involves a cosmological scalar field -- the cosmon, or quintessence field \cite{Wetterich:1987fm,Peebles:1987ek,Caldwell:1997ii,Wetterich:1994bg} (see also \cite{Peccei:1987mm}). The mass of the cosmon decreases with time; for cosmological solutions it turns out to be of the order of the Hubble scale \cite{Wetterich:1994bg}. This makes this field essentially massless as far as black-hole physics is concerned. The cosmon is thus a well-motivated candidate which could influence the physics of black holes.

Cosmon lumps are bound objects, for which the scalar and gravitational fields combine to form non-linear solutions of their field equations. Such solutions have been discussed in detail in \cite{Wetterich:2001cn} for situations where the influence of the cosmon potential is negligible. The salient features are: (i) the lumps are characterized by their mass and a suitably defined ``cosmon charge"; (ii) the black-hole event horizon disappears for all solutions with non-zero cosmon charge; (iii) for small cosmon charge, the solutions closely resemble black-hole solutions when seen from the outside, and the horizon is replaced by a sharp transition to a region of high red-shift. It has been speculated that such lumps, perhaps also in presence of an additional matter component, could be regarded as a candidate for dark matter, if sufficiently small objects are produced sufficiently early in cosmology and with sufficient abundance \cite{Wetterich:2001yw}. Indeed, interesting solutions for a cosmon coupled to dark matter \cite{Brouzakis:2005cj} or to neutrinos \cite{Brouzakis:2007aq} have been obtained.

In this Letter we address another issue, namely the singularity at the center of a black hole. For this reason we include the effects of the cosmon potential, which adds further non-linearities to the system of field equations. We find static, spherically symmetric solutions that look from the outside very similar to a black hole, but are completely regular.

\section{Field equations}

We consider the system of a scalar cosmon (or quintessence) field $\varphi$ coupled to gravity. The action reads
\begin{equation}
   S = \int d^4x\,\sqrt{-g} \left[ \frac{M^2 R}{2} 
   - \frac{g^{\mu\nu}}{2}\,\partial_\mu\varphi\,\partial_\nu\varphi
   - V(\varphi) \right] ,
\end{equation}
where $M=1/\sqrt{8\pi G}$ is the reduced Planck mass. In order to investigate static, spherically symmetric solutions we employ the metric $ds^2=-B(r)\,dt^2+A(r)\,dr^2+r^2(d\theta^2+\sin^2\theta\,d\phi^2)$. The field equations $R_{\mu\nu}-\frac12 g_{\mu\nu} R=T_{\mu \nu}/M^2$ will be solved numerically for a given scalar potential $V (\varphi)$. The energy-momentum tensor is diagonal, with energy density $\rho=-T^0{}_0$ and pressures $p_r=T^r{}_r$, $p_\Omega=T^\theta{}_\theta=T^\phi{}_\phi$ given by
\begin{equation}
   \rho = - p_\Omega = \frac{\varphi'^2}{2A} + V(\varphi) \,,
    \qquad
   p_r = \frac{\varphi'^2}{2A} - V(\varphi) \,.
\end{equation}
We shall refer to these static solutions as ``cosmon lumps".

The ten Einstein equations reduce to a set of three ordinary differential equations for the functions $A(r)$, $B(r)$, and $\varphi(r)$. Those equations are supplemented by the Klein-Gordon equation for the scalar field, $\Box\,\varphi=\frac{dV(\varphi)}{d\varphi}$, or explicitly
\begin{equation}\label{KG}
   \frac{1}{A} \left[ \varphi''
    + \left( \frac{2}{r} + \frac{B'}{2B} - \frac{A'}{2A} \right) 
    \varphi' \right] 
   = \frac{dV(\varphi)}{d\varphi} \,.
\end{equation}
These four equations are redundant, and it suffices to use two linear combinations of the Einstein equations in addition to (\ref{KG}):
\begin{equation}
\begin{aligned}
   \frac{1}{r^2} - \frac{1}{r^2 A} + \frac{A'}{r A^2}
   &= \frac{1}{M^2} \left[ \frac{\varphi^{\prime 2}}{2A}
    + V(\varphi) \right] , \\
   \frac{1}{r} \left( \frac{A'}{A} + \frac{B'}{B} \right)
   &= \frac{\varphi^{\prime 2}}{M^2} \,.
\end{aligned}
\end{equation}

\section{Gray holes}

Time-translation symmetry gives rise to a conserved current $J^\mu=K_\lambda R^{\lambda\mu}$ built using the time-like Killing vector $K^\lambda= (1,0,0,0)$. The corresponding conserved charge can be identified with the total gravitating mass of the system,
\begin{equation}
   m = 2 M^2\!\int\!d^3x\,\sqrt{g_3}\,n_\mu J^\mu 
   = 4\pi M^2\lim_{r\to\infty} \frac{r^2 B'(r)}{\sqrt{A(r) B(r)}} \,.
\end{equation}
Here $g_3$ is the determinant of the induced three-dimensional metric, and $n_\mu=(-\sqrt{B},0,0,0)$ is a time-like normal unit vector. Here and below, a prime denotes a derivative with respect to $r$. Generalizing this result, the enclosed gravitating mass inside a sphere with radius $r$ can be defined as
\begin{equation}
   m(r) = 4\pi M^2\,\frac{r^2 B'(r)}{\sqrt{A(r) B(r)}} \,.
\end{equation}
The strength of the cosmon field can be measured by the enclosed ``cosmon charge"
\begin{equation}
   q(r)\equiv 4\pi M^2 r^2\,\sqrt{\frac{B(r)}{A(r)}}\,
   \hat\varphi'(r) \,,
\end{equation}
where we have introduced the dimensionless scalar field $\hat\varphi(r)=\frac{\sqrt2}{M}\,\varphi(r)$. Note that this charge is not conserved in general, since it is not derived from a global symmetry. By taking appropriate combinations of the field equations, the mass and cosmon charge can be expressed in terms of the potential as
\begin{equation}\label{eq:masscosmoncharge}
\begin{aligned}
   m(r) &= m_0 - 8\pi\int_0^r\!ds\,s^2 \sqrt{A(s) B(s)}\, 
    V(\hat\varphi(s)) \,, \\
   q(r) &= q_0 + 8\pi\int_0^r\!ds\,s^2 \sqrt{A(s) B(s)}\,
    \frac{dV}{d\hat\varphi}(\hat\varphi(s)) \,,
\end{aligned}
\end{equation}
where $m_0$ and $q_0$ are possible contributions localized at the origin. Note that for a free scalar field, i.e., in the absence of a potential, the quantities $m$ and $q$ are independent of $r$. The corresponding solutions are singular at the origin. Specifically, the scalar field $\varphi$ and the Ricci scalar (along with other curvature invariants) diverge for $r\to 0$. For the regular solutions studied in this Letter, on the other hand, the constants $m_0$ and $q_0$ vanish, and the mass and cosmon charge are generated in the region where the potential is non-zero.

A minimally coupled, free scalar field is one of the simplest generalizations of a Schwarzschild black hole. The exact solutions to the Einstein-Klein-Gordon equations with $V=0$ were discussed a long time ago in \cite{Buchdahl:1959nk,Wyman:1981bd} (similar solutions in the context of Brans-Dicke theory were studied in the original paper \cite{Brans:1961sx}). They exhibit a naked singularity at the origin. The solutions are completely specified in terms of $m$, $q$ and the cosmological value $\hat\varphi_\infty$ of the scalar field far away from any sources. Following \cite{Wetterich:2001cn}, we introduce a new radial coordinate $\chi$ by 
\begin{equation}
   r^2 = R_H^2\,\chi^{1-\delta} (1+\chi)^{1+\delta}
\end{equation}
and write the analytical solutions in the form
\begin{equation}\label{eq:freesolutions}
\begin{aligned}
   A(\chi) &= \frac{4\chi(1+\chi)}{(2\chi+1-\delta)^2} \,, \\
   B(\chi) &= \left( \frac{\chi}{1+\chi} \right)^\delta , \\[0.09cm]
   \hat\varphi(\chi) &= \hat\varphi_\infty + \gamma\,\ln B(\chi) \,.
\end{aligned}
\end{equation}
A useful relation is $dr/d\chi=R_H/\sqrt{AB}$. We define the ``would-be horizon" $R_S=2Gm$ and a related scale $R_H=2G\sqrt{m^2+q^2}$, both of which are determined by the mass and cosmon charge of the solution. The coefficients $\gamma$ and $\delta$ are given as
\begin{equation}
   \gamma = \frac{q}{m} \,, \qquad
   \delta = \frac{R_S}{R_H} = \frac{m}{\sqrt{m^2+q^2}} \,.
\end{equation}
The special case $\delta=1$ (or $\gamma=0$) leads to the Schwarzschild solution. All other solutions with $\gamma\ne 0$ have no event horizon. The black hole has become ``gray". 

At large distances ($r\gg R_S$), the solutions for the metric functions obey
\begin{equation}\label{solu}
\begin{aligned}
   A(r) &= \left( 1 - \frac{R_S}{r} \right)^{-1} 
   && \hspace{-5mm}{}- \gamma^2
    \left( \frac{R_S^2}{4r^2} + \dots \right) , \\
   B(r) &= \;\;\, 1 - \frac{R_S}{r} 
   && \hspace{-5mm}{}+ \gamma^2 \left( \frac{R_S^3}{24 r^3}
    + \dots \right) .
\end{aligned}
\end{equation}
Large grey holes are essentially indistinguishable from black holes for an outside observer. This may be demonstrated by assuming that the massive black hole at the center of our galaxy with $m =(3.7\pm 0.4)\cdot 10^6$ solar masses is a cosmon lump with $\gamma=q/m$ of order unity. The star S0-16 approached the center of the black hole up to a distance of about $600 R_S$ \cite{Zucker:2005tz}. In this case, the modifications of the metric coefficients $A$ and $B$ with respect to the Schwarzschild metric are about $7\cdot 10^{-7}\,\gamma^2$ and $2\cdot 10^{-10}\,\gamma^2$, respectively. Obviously, only very large values of $\gamma$ would give rise to noticeable modifications of the metric at large distances.

\section{Regular solutions}

The solutions for a free scalar field explored in the previous section exhibit a naked singularity at $r=0$. However, regular solutions exist if we allow for a non-zero potential for the scalar field. For regular solutions we use Taylor expansions of the form $A(r)=\sum_{n=0}^\infty A_n\,r^n$ for all fields, and find after a straightforward analysis of the field equations
\begin{equation}
\begin{aligned}
   A(r) &= 1 + \frac{V(\hat\varphi_0)}{3M^2}\,r^2 + \dots \,, \\
   B(r) &= B_0 \left( 1 - \frac{V(\hat\varphi_0)}{3M^2}\,r^2
    + \dots \right) , \\
   \hat\varphi(r) &= \hat\varphi_0 + \frac{1}{3M^2}\, 
    \frac{dV(\hat\varphi_0)}{d\hat\varphi}\,r^2 + \dots \,.
\end{aligned}
\end{equation}
Note that the linear coefficients $A_1$, $B_1$, $\hat\varphi_1$ vanish, while $A_0=1$ is fixed by the demand for the absence of a conical singularity. Since the field equations only depend on the ratio $B'/B$, the value of $B_0$ remains undetermined. It must be set in such a way that the metric is asymptotically flat, $B(r)\to 1$ for $r\to\infty$. Thus, for a given potential, the value of $\hat\varphi_0$ is the only free boundary condition.

\begin{figure}
\includegraphics[width=0.9\columnwidth]{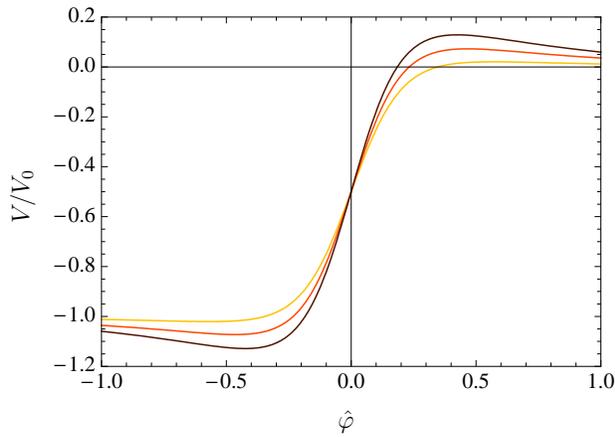}
\caption{\label{fig:potential}
The cosmon potential (\ref{bump}) for $\eta=5$, $\kappa=1$, and $\epsilon=0.5$ (dark), 0.3 (medium), 0.1 (light).}	
\end{figure}

An important observation following from the above relations is that, as $r$ increases away from the origin, the scalar field {\em climbs up\/} the potential, unless $dV(\hat\varphi_0)/d\hat\varphi=0$, which would require a fine-tuning. Since we should require that for $r\to\infty$ the cosmon is driven to a region with zero (or very small positive) potential, there are essentially only two types of potentials that could give rise to regular solutions: ``slope" potentials, where the scalar potential starts out at some negative value at the center of the cosmon lump and monotonically increases toward zero (or a very small positive value) for large $r$, or ``bump" potentials, where between an initial negative value and an asymptotic cosmological value the cosmon field climbs over a local maximum of the potential. From (\ref{eq:masscosmoncharge}) it follows that slope potentials always give rise to positive mass and cosmon charge. The same is true for bump potentials provided the potential is sufficiently negative inside the cosmon lump. For phenomenological purposes bump potentials appear more attractive, as they provide a potential barrier against collapse of the scalar field into the true ground state (with negative potential). Also, in this case it is possible to achieve that the potential asymptotically approaches the exponential form $V(\hat\varphi)\propto e^{-2\kappa\hat\varphi}$, which is a particularly natural potential for the cosmon \cite{Wetterich:2008sx}.

A simple model for a cosmon potential with a bump is 
\begin{equation}\label{bump}
   V(\hat\varphi) = - V_0 \left[ (1+\epsilon)\,
   \frac{1 - \tanh\eta\hat\varphi}{2}
   - \epsilon\,\frac{1 - \tanh\kappa\hat\varphi}{2} \right] 
\end{equation}
with positive parameters $V_0$ (typically of order $M^4$), $\epsilon\ll 1$, and $\kappa<\eta$. It starts at $-V_0$ for $\hat\varphi\ll 0$ and approaches zero from above for $\hat\varphi\gg 0$, i.e.\ $V(\hat\varphi)\approx V_0\,\epsilon\,e^{-2\kappa\hat\varphi}$, such that this model may explain the dark-energy content of the universe on large scales. The limit $\epsilon\to 0$ leads to a simple slope potential. Figure~\ref{fig:potential} shows the model potential for some parameter choices.

Our goal is to find regular cosmon-lump solutions, for which at $r=0$ the scalar potential takes a negative value  somewhere to the right of the minimum of the potential, while asymptotically (for $r\to\infty$) it approaches a small positive value to the right of the maximum. The radial dependence of the potential within a cosmon lump introduces another length scale $R_C$ besides the Schwarzschild radius. It is the scale beyond which the effects of the potential can be neglected, so that the mass and cosmon charge computed from (\ref{eq:masscosmoncharge}) remain constant as $r$ is further increased. For $r>R_C$ the solutions thus converge toward the solutions (\ref{solu}) obtained for a free scalar field. In practice, we solve the field equations numerically for a chosen value of $\hat\varphi_0$ and compute the parameter $\gamma$ at a sufficiently large value of the radial coordinate ($r>R_C$). We then match onto the free solution (\ref{solu}) and determine the coefficient $B_0$, which fixes the absolute values $m$ and $q$ of the solution.  

\begin{figure}
\includegraphics[width=0.86\columnwidth]{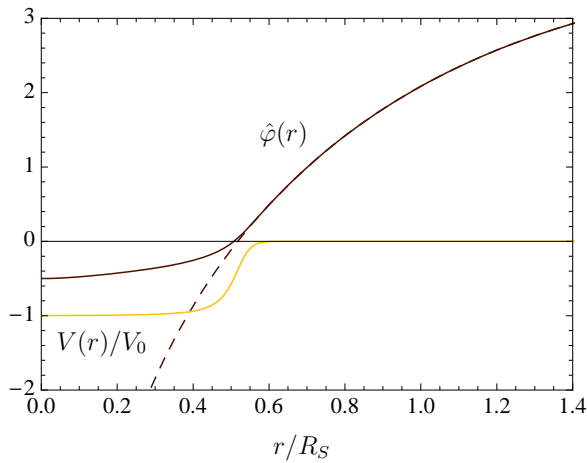}
\caption{\label{fig:field_potential}
The scalar field $\hat\varphi$ (dark) and the potential $V/V_0$ (light) as functions of $r$. The dashed lines show the scalar field obtained with $V=0$, which diverges for $r\to 0$.}
\end{figure}

\begin{figure}
\includegraphics[width=0.86\columnwidth]{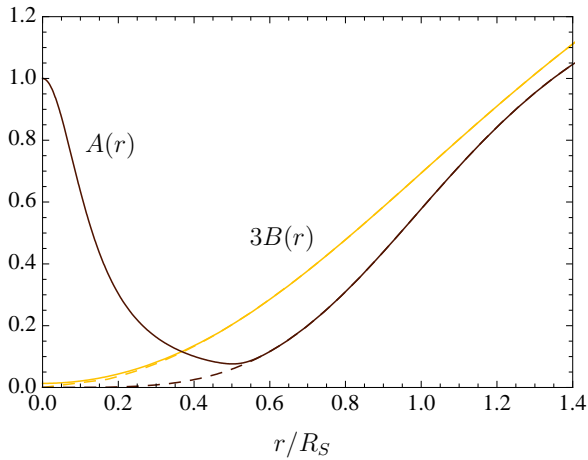}
\caption{\label{fig:metric}
Metric coefficients $A$ (dark) and $3B$ (light) as functions of $r$. Solid lines show the regular cosmon-lump solutions obtained with the potential (\ref{bump}), while dashed lines show the singular solutions obtained with $V=0$.}
\end{figure}

As a particular example we discuss the solution obtained with the potential parameters $\eta=5$, $\kappa=3$, $\epsilon=0.1$ and the starting value $\hat\varphi_0=-0.5$ for the scalar field at the center of the cosmon lump. The corresponding asymptotic values of the mass and cosmon charge are $m\approx 165$ and $q\approx 331$ in units of $M^3/\sqrt{V_0}$, which implies $\gamma\approx 1.79$. Figure~\ref{fig:field_potential} shows the radial profile of the scalar field and of the potential as functions of $r$ in units of the Schwarzschild radius $R_S=2Gm$. Note that the potential becomes negligible for $r>R_C$ with $R_C\approx 0.55 R_S$. The dashed line shows for comparison the solution obtained for a free scalar field, which is singular at $r=0$. The presence of the potential removes the naked singularity and gives rise to solutions that are regular everywhere. The corresponding results for the metric functions $A$ and $B$ are shown in Figure~\ref{fig:metric}. Once again the dashed lines correspond to the solutions without the potential. They are indistinguishable from the exact ones for $r/R_S>0.55$. 

\begin{figure}
\includegraphics[width=0.9\columnwidth]{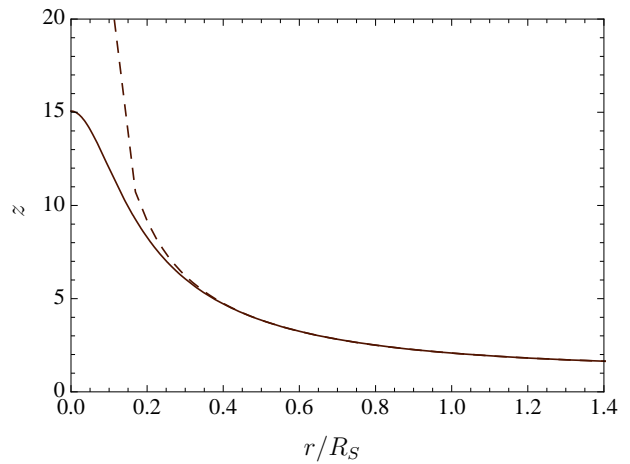}
\caption{\label{fig:redshift}
The gravitational redshift $z(r)$ for the cases of a free ($V=0$, dashed) and interacting ($V\ne 0$, solid) scalar field. In the first case, the redshift diverges at the singularity. In the latter case, the redshift is finite everywhere but becomes large at the center.}
\end{figure}

At large distances ($r>\mbox{few}\,R_S$) the metric closely resembles the Schwarzschild solution, but it is strongly modified around and inside the Schwarzschild scale $R_S$. We find no horizon, i.e.\ no coordinate singularity, and no physical singularity at the center. Note, in particular, that $B(r)$ converges to a non-zero value $B_0\approx 0.0044$ for $r\to 0$. In the Schwarzschild metric, the gravitational redshift of a resting source
\begin{equation}\label{eq:redshift}
   z(r) = \sqrt{\frac{1}{B(r)}} - 1
\end{equation}
diverges at $R_S$. Our solutions feature no redshift horizon, but a large redshift value at the center of the cosmon lump, see Figure~\ref{fig:redshift}. This behavior might simulate an event horizon, if an observer does not measure precisely enough. By choice of the initial parameters the redshift at the origin can be made arbitrarily large, but it is always finite. 

The values of the mass $m$ and cosmon charge $q$ of the solution depend strongly on the initial value $\hat\varphi_0$ and on the overall scale $V_0$ of the potential. In this way, a wide range of mass scales can be obtained in our model. Interestingly, the value of $\gamma$ asymptotically reaches $\sqrt 3$ in the large mass limit, $m>10^3\,M^3/\sqrt{V_0}$.

\section{Stability analysis}

In order to estimate the stability of the regular solutions, we introduce radial perturbations of the metric and the scalar field, which are allowed to evolve in time. The full time-dependent field equations are then linearized about the static solutions (denoted by a subscript $s$), using the ansatz
\begin{equation}\begin{aligned}
   \hat\varphi(t,r) &= \hat\varphi_s(r) + \delta\hat\varphi(t,r)
    \,, \\
   A(t,r) &= A_s(r) + \delta A(t,r) \,, \\
   B(t,r) &= B_s(r) + \delta B(t,r) \,.
\end{aligned}
\end{equation}
Only terms of first order in the perturbations and their derivatives are taken into account. In contrast to the three static equations, one finds four independent equations for the time dependence of the fields, since $R_{01}$ and $T_{01}=\dot\varphi\varphi'$ (the radial energy-flux density) do not vanish. Following \cite{Jetzer:1992ag}, the four linearized field equations can be combined to obtain one single perturbation equation for $\delta\hat\varphi$. We look for solutions of this equation of the form
\begin{equation}
   \delta\hat\varphi(t,r) = e^{i\sigma t}\,\frac{\Psi(r)}{r} \,.
\end{equation}
If $\sigma$ is real, the perturbation is a small oscillation about the static solution. However, imaginary $\sigma$ leads to a dynamical instability. 

\begin{figure}
\includegraphics[width=0.9\columnwidth]{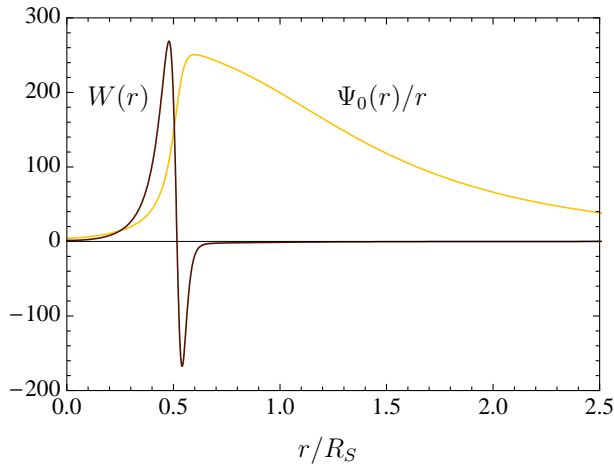}
\caption{\label{fig:Veff}
The pulsation potential $W(r)$ (dark, in units of $R_S^{-2}$) features a distinctive minimum in the vicinity of $R_C$, which supports a shallow bound state. The corresponding eigenfunction $\Psi_0(r)/r$ (light, in arbitrary units) determines the radial profile of the unstable perturbation to the cosmon lump.}
\end{figure}

Introducing a new radial coordinate $\varrho(r)$ obeying $\varrho(0)=0$ and $d\varrho/dr=\sqrt{A_s/B_s}$, the perturbation equation for $\delta\hat\varphi$ implies the ``pulsation equation"
\begin{equation}\label{eq:pulsationequation}
   - \frac{d^2\Psi}{d\varrho^2} + W\,\Psi = \sigma^2\,\Psi
\end{equation}
with the ``pulsation potential''
\begin{eqnarray}
   W &=& \frac{B_s}{A_s} \Bigg[ \frac{1}{2r} \left(
    \frac{B'_s}{B_s} - \frac{A'_s}{A_s} \right)
    - \frac{r\hat\varphi'^2_s}{4}
    \left( \frac{2}{r} + \frac{B'_s}{B_s} - \frac{A'_s}{A_s} \right)
    \nonumber\\
   &&\mbox{}+ \frac{2r\hat\varphi'_s A_s}{M^2}\,
    \frac{dV(\hat\varphi_s)}{d\hat\varphi}
    + \frac{2A_s}{M^2}\,\frac{d^2V(\hat\varphi_s)}{d\hat\varphi^2}
    \Bigg] \,.
\end{eqnarray}
The asymptotic behavior of this function is
\begin{equation}
\begin{aligned}
   W(r\to 0) &= \frac{2B_0}{M^2} \left[ 
    \frac{d^2V(\hat\varphi_0)}{d\hat\varphi^2}
    - \frac{V(\hat\varphi_0)}{3} \right] + O(r^2) , \\
   W(r\to\infty) &= \frac{R_S}{r^3} + O(r^{-4}) \,.
\end{aligned}
\end{equation}
The dark curve in Figure~\ref{fig:Veff} shows the pulsation potential for the parameter choices considered in the previous section. 

Equation~(\ref{eq:pulsationequation}) resembles the one-dimensional Schr\"odinger equation for a particle in the potential $W$. The eigenvalues $\sigma^2$ can be calculated with the boundary conditions that $\Psi(0)=0$, so that $\delta\hat\varphi$ remains regular at the origin, and $\Psi(\infty)=0$, such that the wave function can be normalized. The knowledge of the lowest eigenvalue $\sigma_0^2$ is sufficient for our purposes, since it determines the time evolution of the perturbations. The cosmon-lump solution is instable if a bound state exists, i.e., a solution to the pulsation equation with negative eigenvalue $\sigma_0^2<0$. For our particular model, we find that the negative spike of the potential, centered near $R_C\approx 0.55 R_S$, supports a very shallow bound state, whose eigenvalue $\sigma_0^2\approx -0.197 R_S^{-2}$ is much smaller in magnitude than the minimum of the potential ($W_{\rm min}\approx -167 R_S^{-2}$). The radial profile $\Psi_0(r)/r$ corresponding to this eigenvalue is shown by the light curve in the figure. 

We have explored a large set of parameter choices for the cosmon potential (\ref{bump}) as well as different initial values $\hat\varphi_0$ for the scalar field at the center of the cosmon lump, and in all cases a negative eigenvalue $\sigma_0^2$ exists. We thus conclude that the regular, static solutions to the Einstein-Klein-Gordon equations obtained in this Letter using a non-trivial scalar potential, which {\em a priori\/} are not in conflict with the cosmic censorship conjecture \cite{Penrose:1969pc}, are dynamically unstable. In this respect they share the fate of the known singular solutions explored in the literature \cite{Jetzer:1992ag,Liebling:1996dx,Choptuik:1997rt}.

\section{Conclusions}

We have shown that a rather simple system of a self-interacting scalar field minimally coupled to gravity allows for static, spherically symmetric solutions to the Einstein-Klein-Gordon equations that for an outside observer look very close to the black-hole solutions in pure gravity, but have neither a horizon nor a central (naked) singularity. Such regular solutions are of high conceptual interest. For the choice of the scalar potential considered in this Letter, the static solutions turn out to be dynamically unstable. Unfortunately, they can therefore not describe astrophysical objects such as the black holes in the center of galaxies or dark-matter lumps. 

Our analysis leaves open the questions whether similar solutions exist, which are both regular and stable. Perhaps such solutions can be found by modifying the cosmon potential, or by adding a matter component coupled to the cosmon field. If such regular ``gray holes" without event horizons exist, this would revolutionize the conventional view of black holes and related puzzles such as the information paradox.


\begin{thebibliography}{99}

\bibitem{Buchdahl:1959nk}
  H.~A.~Buchdahl,
  Phys.\ Rev.\  {\bf 115}, 1325 (1959).

\bibitem{Brans:1961sx}
  C.~Brans and R.~H.~Dicke,
  Phys.\ Rev.\  {\bf 124}, 925 (1961).

\bibitem{Wyman:1981bd}
  M.~Wyman,
  Phys.\ Rev.\  D {\bf 24} (1981) 839.

\bibitem{Wetterich:2001cn}
  C.~Wetterich,
  Phys.\ Lett.\  B {\bf 522}, 5 (2001)
  [arXiv:astro-ph/0108411].

\bibitem{vanPutten:1996mt}
  M.~H.~P.~van Putten,
  Phys.\ Rev.\  D {\bf 54}, 5931 (1996)
  [arXiv:gr-qc/9607074].

\bibitem{Jetzer:1992ag}
  P.~Jetzer and D.~Scialom,
  Phys.\ Lett.\  A {\bf 169}, 12 (1992).

\bibitem{Liebling:1996dx}
  S.~L.~Liebling and M.~W.~Choptuik,
  Phys.\ Rev.\ Lett.\  {\bf 77}, 1424 (1996)
  [arXiv:gr-qc/9606057].

\bibitem{Choptuik:1997rt}
  M.~W.~Choptuik, E.~W.~Hirschmann and S.~L.~Liebling,
  Phys.\ Rev.\  D {\bf 55}, 6014 (1997)
  [arXiv:gr-qc/9701011].

\bibitem{Wetterich:1987fm}
  C.~Wetterich,
  Nucl.\ Phys.\  B {\bf 302}, 668 (1988).

\bibitem{Peebles:1987ek}
  P.~J.~E.~Peebles and B.~Ratra,
  Astrophys.\ J.\  {\bf 325}, L17 (1988);
%
  B.~Ratra and P.~J.~E.~Peebles,
  Phys.\ Rev.\  D {\bf 37}, 3406 (1988).

\bibitem{Caldwell:1997ii}
  R.~R.~Caldwell, R.~Dave and P.~J.~Steinhardt,
  Phys.\ Rev.\ Lett.\  {\bf 80}, 1582 (1998)
  [arXiv:astro-ph/9708069].

\bibitem{Wetterich:1994bg}
  C.~Wetterich,
  Astron.\ Astrophys.\  {\bf 301}, 321 (1995)
  [arXiv:hep-th/9408025].

\bibitem{Peccei:1987mm}
  R.~D.~Peccei, J.~Sola and C.~Wetterich,
  Phys.\ Lett.\  B {\bf 195}, 183 (1987).

\bibitem{Wetterich:2001yw}
  C.~Wetterich,
  Phys.\ Rev.\  D {\bf 65}, 123512 (2002)
  [arXiv:hep-ph/0108266].

\bibitem{Brouzakis:2005cj}
  N.~Brouzakis and N.~Tetradis,
  JCAP {\bf 0601}, 004 (2006)
  [arXiv:astro-ph/0509755].

\bibitem{Brouzakis:2007aq}
  N.~Brouzakis, N.~Tetradis and C.~Wetterich,
  arXiv:0711.2226 [astro-ph].

\bibitem{Zucker:2005tz}
  S.~Zucker, T.~Alexander, S.~Gillessen, F.~Eisenhauer and R.~Genzel,
  Astrophys.\ J.\  {\bf 639}, L21 (2006)
  [arXiv:astro-ph/0509105]; 
%
for a review, see:
  T.~Alexander,
  Phys.\ Rept.\  {\bf 419}, 65 (2005)
  [arXiv:astro-ph/0508106].

\bibitem{Wetterich:2008sx}
  C.~Wetterich,
  arXiv:0801.3208 [hep-th].

\bibitem{Penrose:1969pc}
  R.~Penrose,
  Riv.\ Nuovo Cim.\  {\bf 1}, 252 (1969)
  [Gen.\ Rel.\ Grav.\  {\bf 34}, 1141 (2002)];
%
for a review, see:
  R.~M.~Wald,
  arXiv:gr-qc/9710068.

\end{thebibliography}
\end{document}